\documentclass[apj]{emulateapj}
\usepackage{epsfig}

\begin{document}
\title{SN2007ax : An Extremely Faint Type Ia Supernova}

\author{M. M. Kasliwal\altaffilmark{1,2},
E. O. Ofek\altaffilmark{1}, A. Gal-Yam\altaffilmark{1}, A. Rau\altaffilmark{1}, 
P. J. Brown\altaffilmark{3}, S. B. Cenko\altaffilmark{4}, P. B. Cameron\altaffilmark{1},
R. Quimby\altaffilmark{1}, S. R. Kulkarni\altaffilmark{1},  L. Bildsten\altaffilmark{5}, P. Milne\altaffilmark{6}}

\altaffiltext{1}{Astronomy Department, California Institute of Technology, 105-24, Pasadena, CA 91125, USA}
\altaffiltext{2}{George Ellory Hale Fellow, Gordon and Betty Moore Foundation}
\altaffiltext{3}{Department of Astronomy and Astrophysics, Pennsylvania State University, 
525 Davey Laboratory, University Park, PA 16802}
\altaffiltext{4}{ Space Radiation Laboratory, California Institute of Technology, MS 220-47, Pasadena, CA 91125}
\altaffiltext{5}{Kavli Institute for Theoretical Physics and Department of Physics, Kohn Hall, 
University of California, Santa Barbara, CA 93106}
\altaffiltext{6}{Steward Observatory, 933 N. Cherry Ave., Tucson, AZ 85721} 
\email{mansi@astro.caltech.edu}

\begin{abstract}

We present multi-band photometric and optical spectroscopic observations 
of SN2007ax, the faintest and reddest Type Ia supernova (SN\,Ia) yet observed. With 
M$_{\rm B}$ = $-$15.9 and ($B-V$)$_{\rm max}$ = 1.2, this SN is over half a magnitude fainter
at maximum light than any other SN\,Ia. Similar to subluminous SN2005ke, SN2007ax also appears to show excess in UV emission 
at late time. Traditionally, $\Delta$m$_{15}$($B$) has been used to parameterize the decline rate 
for SNe\,Ia. However, the B-band transition from fast to slow decline occurs sooner than 
15 days for faint SNe\,Ia. Therefore we suggest that a more physically motivated parameter, the
time of intersection of the two slopes, be used instead. Only by explaining the faintest (and the brightest) 
supernovae, we can thoroughly understand the physics of thermonuclear explosions. We suggest that future 
surveys should carefully design their cadence, depth, pointings and follow-up to find an unbiased 
sample of extremely faint members of this subclass of faint SNe\,Ia.    

\end{abstract}


\section{Introduction}

Inspired by the application as a standard cosmological candle, the 
progress in understanding Type Ia supernovae (SNe\,Ia) has grown in
leaps and bounds. However, the understanding of their weakest 
subluminous cousins has been purposefully overlooked as their
atypical light curve and atypical spectra make them contaminants for 
cosmological studies. We suggest here some characteristics that make the 
physics of the explosions of faint SNe\,Ia intriguing in their own right. 

In this paper, we present SN2007ax which, with a peak absolute magnitude of 
M$_{\rm B}$ = $-$15.9 and ($B-V$)$_{\rm max}$ = 1.2, 
is the faintest and reddest Type Ia supernova yet discovered. Although the class
of SNe\,Ia is remarkably homogenous, subluminous SNe\,Ia show atypical spectral 
and light curve features (\citealt{g+04}, \citealt{t+07}). Photometrically, not only do they fade much 
faster than predicted by the Phillips relation, they are also very red 
at maximum and (at least SN2005ke and SN2007ax) appear to show UV excess at late-time. 
Spectroscopically, they have broad \ion{Ti}{2} features and moderate expansion velocities.

SN2007ax was discovered in NGC\,2577 on UT 2007 Mar 21.978 by \citet{a07}
at an unfiltered magnitude of 17.2. Upper limits of $>$18.5 mag 
on Mar 17.636 and $>$19.0 mag on Mar 9.959 were also reported.
Spectra obtained on Mar 26 by \citet{bm+07} and \citet{mf07} 
showed that it was a SN\,Ia near maximum light similar 
to SN1991bg. 

In this paper, we present multi-epoch, multi-band imaging and 
spectroscopic follow up of SN2007ax including optical, ultraviolet,  
and near-infrared. We summarize our observations in \S\,2, present 
our analysis and comparison with other faint SNe\,Ia in \S\,3 
and discuss possible scenarios for faint thermonuclear explosions in \S\,4. 
We conclude with how future surveys can 
systematically design their cadence, limiting magnitude and pointings to search for 
more members belonging to this subclass of faint SNe\,Ia. 

\section{Observations and Data Reduction}

The automated Palomar 60-inch telescope \citep{cf+06} started daily observations 
of SN2007ax on UT 2007 Mar 29 in $g^{\prime}$ and $r^{\prime}$ bands .
Data were reduced using custom routines. Aperture photometry was
done after image subtraction using two custom modifications of 
the ISIS algorithm \citep{al98}, {\it hotpants}\footnote{http://www.astro.washington.edu/becker/hotpants.html} 
and {\it mkdifflc} (\citealt{gy+04}, \citealt{gy+07}). The two 
reductions gave consistent results. Errors were estimated by first placing
artificial sources of the same brightness and at the same distance from the galaxy
center as the SN and then measuring the scatter in measured magnitudes. 
Finally, the zeropoint was calibrated with reference magnitudes of stars from the 
Sloan Digital Sky Survey \citep{a+07}.

We triggered Target of Opportunity observations to obtain spectra with
the Double Beam Spectrograph \citep{og82} on the Hale 200-inch telecope. Two spectra
were obtained around maximum light (UT 2007 Mar 29 and Mar 30) and a third a 
fortnight later (Apr 13). Spectra were taken using the red grating 158/7500, 
blue grating 300/3990 and using a dichroic to split the light at 5500\,\AA\ . This 
gave us a total wavelength
coverage of 3800\,\AA\ - 9000\,\AA\ and dispersion of 4.9\,\AA\ pix$^{-1}$ and 
2.1\,\AA\ pix$^{-1}$ on the red and blue side, respectively. Data were reduced 
using the standard IRAF\footnote{IRAF is distributed by the National Optical Astronomy 
Observatories, which are operated by the Association of Universities for Research
in Astronomy, Inc., under cooperative agreement with the National
Science Foundation} package {\it apall}.   

We triggered Swift Target of Opportunity observations for SN2007ax 
starting UT 2007 Mar 29.84 and obtained eight epochs of roughly five
kiloseconds each distributed between the $uvw2, uvm2, uvw1,
u, b$ and  $v$ bands. We also obtained a reference
image over eight months after peak to subtract galaxy light. Aperture 
photometry was performed using a 3$\arcsec$ circular radius. To estimate the galaxy 
brightness at this location, a 3$\arcsec$ aperture at the supernova
position in the reference image was used. \citet{p+07} photometric zeropoints were
applied after appropriately scaling for aperture size. For consistency 
with calibration, a 5$\arcsec$ aperture was used in the computation of coincidence 
loss. The supernova is detected in $uvw1$ in four epochs, and not detected in the $uvw2$ 
and $uvm2$ filters. The $b$ band light curve was independently reduced
using image subtraction with consistent results. We note that due to the faintness  
of the supernova and brightness of galaxy background, coincidence loss is dominated by 
the galaxy light and not a point source, possibly introducing a systematic error in the
Swift $u$, $b$, and $v$ bands.

Further late-time $BVRI$ observations were obtained using the SLOTIS and Bok 
telescopes and light curves were obtained using image subtraction based on ISIS
and IRAF routines. We also obtained near infrared $K^{\prime}$ imaging using the Keck 
NIRC2 instrument with Natural Guide Star adaptive optics on UT 2007 Apr 4.

\section{Analysis}

We present analysis of the optical and ultra-violet light curve and optical
spectrum of SN2007ax below. We also compare it to other subluminous SN Ia. 
We adopt a distance modulus of 32.2 (B. Tully, pers. comm.) to NGC\,2577.

\subsection{Optical Light Curve}

\begin{figure}
\epsscale{0.5}
\centerline{\includegraphics[angle=90, width=3.5in]{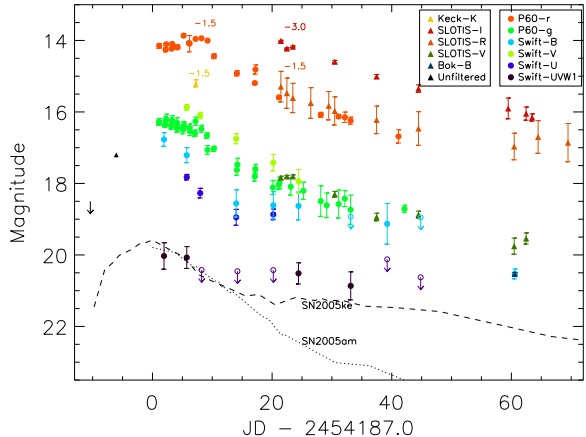}}
\caption{ Multi-band light curve of SN2007ax based on data from P60, 
Swift/UVOT, SLOTIS, Bok and Keck II/NIRC2. Unfiltered magnitudes from 
\citet{a07}. Note that similar to subluminous SN2005ke
(dashed line), SN2007ax also appears to show an excess in UV emission at 
t$>$20days while typical SNe\,Ia (SN2005am, dotted line) continue to decline.  
\label{fig:lc}}
\end{figure}

We plot the multi-band light curve of SN2007ax in Figure~\ref{fig:lc}.
The key characteristic of SN2007ax is its rapid decline. Traditionally,
$\Delta$m$_{15}$ (the difference between the peak B-mag and  the B-mag 
15 days after the peak) has been used to parametrize the decline of the light curve. 
However, this parameter can be misleading when applied
to the faint SNe\,Ia because the knee in their light curve (transition 
from fast initial decline to slow late-time decline) is sooner than fifteen days from
the peak. Therefore, we choose to compare the light curves of subluminous Ia using
three parameters first introduced by \citet{p84}  - initial slope ($\beta$), 
late-time slope ($\gamma$) and the time of intersection of the two slopes (t$_{b}$). 
This time of intersection parameter (defined from maximum in B-mag) was also used 
by \citet{hp+96} as t$_{2}^{\rm B}$ and shown 
to be empirically proportional to $\Delta$m$_{15}$ for some SNe\,Ia. 

\begin{figure}
\epsscale{0.5}
\centerline{\includegraphics[angle=90, width=3.5in]{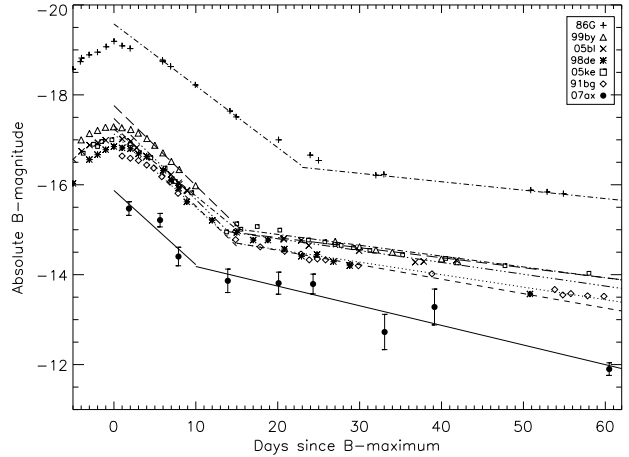}}
\caption{B-band light curve of SN2007ax in comparison with other subluminous
SNe\,Ia. The best linear fits are overplotted and give the early-time and 
late-time slope. We note that the time of intersection, t$_{b}$ of the early-time and late-time
slopes is more strongly correlated with the absolute magnitude than the slopes, $\alpha$ and $\beta$.}
\label{fig:lczoom}
\end{figure}

\thispagestyle{empty}
{
\begin{table*}
\begin{center}
\caption[]{\bf Comparison of  faint SNe\,Ia}
\begin{tabular}{llccccccl}
\hline
\hline
Supernova & Galaxy & DM & M$_{B,max}$ & $\alpha$ & $\beta$ & t$_{b}$ & ($B-V$)$_{\rm max}$ & Reference \cr
          &        &    &  mag      &  mag/day & mag/day            & days  & mag             &   \cr   
\hline
SN2007ax  & NGC\,2577  & 32.2 & $-$15.9$\pm$0.2 & 0.16 & 0.04 & 10.3 & 1.2 & This Paper   \cr
SN1991bg  & NGC\,4374  & 31.2 & $-$16.6$\pm$0.3 & 0.16 & 0.03 & 14.8 & 0.8 & \citet{l+93},\citet{f+92} \cr
SN1998de  & NGC\,252   & 34.3 & $-$16.8$\pm$0.2 & 0.18 & 0.03 & 14.5 & 0.7 & \citet{m+01} \cr 
SN2005ke  & NGC\,1371  & 31.8 & $-$17.0$\pm$0.2 & 0.15 & 0.02 & 14.9 & 0.7 & \citet{i+06} \cr
SN2005bl  & NGC\,4070  & 35.1 & $-$17.2$\pm$0.2 & 0.18 & .03 & 14.0 & 0.6 & \citet{t+07} \cr
SN1999by  & NGC\,2841  & 30.9 & $-$17.3$\pm$0.2 & 0.18 & 0.02 & 16.0 & 0.5 & \citet{g+04} \cr 
\hline
\hline
\end{tabular}
\end{center}

\label{tab:comparelc}
\end{table*}
}

For the subclass of faint SNe\,Ia, we find that t$_{b}$ is better correlated with the
peak absolute B-mag than $\beta$ and $\gamma$ slopes of the B-band light curve. We fit 
an empirical relation to the intersection time as a function of peak absolute magnitude and find that 
$M_{\rm B} = -13.7 (\pm 0.5)  - 0.22 (\pm 0.03)\times t_{b}$. Moreover, this transition to slower
decline should represent the time at which the optical depth to thermalized
radiation becomes thin. We report these three parameters for a sample of subluminous SNe\,Ia 
in Table~\ref{tab:comparelc} and and show the linear fits in Figure~\ref{fig:lczoom}.

Another crucial property of subluminous SNe\,Ia is that the fainter they are, the redder
they are at maximum. We find that SN2007ax is consistent within uncertainties of 
the empirical relation derived first by \citet{g+04} : 
$M_{\rm B} = -18.7 + (B-V)_{\rm max} \times 2.68 (\pm 0.32)$. 
This relation predicts a color in the range of 1.0--1.3 mag and we observe 1.2 $\pm$ 0.1 mag. 
This color has been derived based on synthetic photometry of the spectra around maximum.

\subsection{Ultraviolet Light Curve}


In Figure~\ref{fig:lc}, we compare the Swift UVOT light curve of 
SN2007ax to another subluminous SN\,Ia 2005ke \citep{i+06} and a typical
SN\,Ia 2005am \citep{br+05}. The key similarity
between SN2005ke and SN2007ax is that both show an excess in UV starting $\approx$20 days
after the peak. \citet{i+06} propose that SN2005ke showed a UV excess due to 
circumstellar interaction. Perhaps, subluminous supernovae are optically thin below 3800\,\AA\ 
simply due to lower production of iron-group elements. The question of 
whether UV excess is a more general property of faint SNe\,Ia merits further 
investigation with timely follow-up of a larger sample. With a larger sample,
one could also consider whether the break in the UV light curve also depends
on absolute magnitude.

\subsection{Spectral Evolution}

\begin{figure}
\epsscale{0.5}
\centerline{\includegraphics[angle=90, width=3.5in]{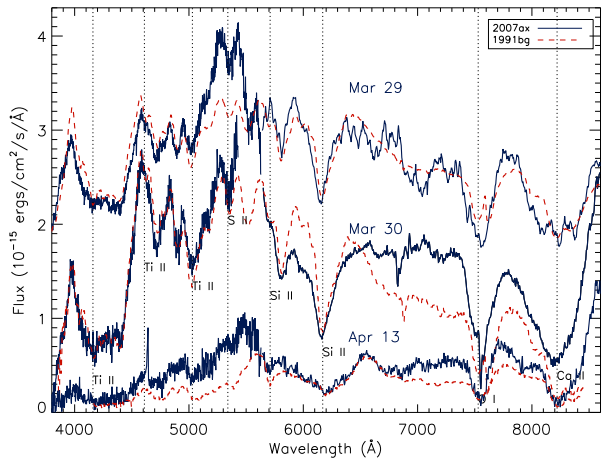}}
\caption{Three epochs of P200 DBSP spectra of SN2007ax (with arbitrary
vertical offsets for clarity). Overplotted is another subluminous Type Ia supernova,
SN1991bg, one day, two days and sixteen days after the peak
(scaled by a multiplicative factor for comparison).}
\label{fig:spec}
\end{figure}

We compare optical spectra of SN2007ax to SN1991bg in Figure~\ref{fig:spec}.
The prominent absorption features are \ion{Ti}{2}, \ion{O}{1}, \ion{Si}{2} and 
\ion{Ca}{1}. The presence of intermediate mass elements like Oxygen and Titanium is 
indicative of the presence of unburned material or a low burning efficiency. The 
absorption features become broader as the supernova evolves.
Comparing our spectra to SN1991bg one day, two days and sixteen days after maximum in
B-band, we find that the spectra are very similar. In the first epoch, we see a hint of carbon
in the small bump immediately redward of the \ion{Si}{2} feature at 6150 \,\AA\ .
However, the signal-to-noise ratio in the spectrum is too low for any conclusive evidence.

Using the technique described by \citet{n+95}, we estimate the temperature
diagnostic R(\ion{Si}{2}) -- the ratio of the depths of the two \ion{Si}{2}
 features at 5800 \,\AA\ and 6150 \,\AA\ -- to be 0.33. This is
smaller than what is implied by the empirical relations derived by \citet{g+04} and 
\citet{t+07}.

We also measure the velocity of the \ion{Si}{2} 6150\,\AA\ line in the two epochs 
around maximum and we obtain 9300 km~s$^{-1}$ and 8800 km~s$^{-1}$. This is 
consistent with lower velocities observed in other faint SNe\,Ia \citep{b+05}.

\subsection{NIR Imaging and Extinction}
We measure a K$^{\prime}$ magnitude of  16.7 $\pm$ 0.1 on UT 2007 Apr 4. We determined the 
contribution of galaxy light at the supernova position by fitting a Sersic
profile to the galaxy using GALFIT \citep{p+02}. The best-fit parameters are : 
a Sersic index of 1.90, axis ratio of 0.60, effective radius of 4.98$\arcsec$, 
position angle of 105.6$^{\circ}$ and diskiness of $-$0.14. We find no evidence of 
dust lanes in this image suggesting that the host extinction is minimal. 
This is also consistent with the absence of the interstellar Na D line at 5893\,\AA\ . 
We compute an upper limit on the equivalent width 
as 0.1\,\AA\ .\ Using the relations derived in \citet{t+03}, we get an 
upper limit of E($B-V$)  $<$ 0.01 mag on the extinction. 

Based on the Galactic position, l=201.1$^{\circ}$, b=29.6$^{\circ}$, the extinction
along the line of sight is E($B-V$)=0.054 mag \citep{s+98}. Therefore, 
we account for A$_{\rm B}$ = 0.23 and A$_{\rm v}$= 0.18 
in our calculations of absolute magnitude and luminosities. 

\subsection{Bolometric Luminosity and $^{56}$Ni Mass}
 
\citet{a85} gives an estimate of the $^{56}$Ni mass in the ejecta using 
the peak bolometric luminosity and the rise time:
$$M_{Ni} = L_{43} \times [6.31~\exp(-t_{r}/8.8) + 1.43~\exp(-t_{r}/111)]^{-1}$$
For SN2007ax, the extinction-corrected peak bolometric luminosity is 
2.3$\times$10$^{42}$ ergs~s$^{-1}$.
We estimate this by using the photometric points to calibrate our spectrum near
maximum light and integrating. 
The rise-time is unknown and unfortunately, the literature somewhat arbitrarily assume 
17 days for faint SNe\,Ia and 19.5 days for typical SNe\,Ia. Recently, 
\citet{t+07} used SN1999by early-time data to estimate a rise-time of 14 days. 
The only observational constraint we have for SN2007ax is that the rise-time is longer than 6 days. 
Thus, for the range of rise-times from 6--14 days, we find a  $^{56}$Ni 
mass of 0.05--0.09 M$_{\odot}$. This is consistent with other techniques to estimate
$^{56}$Ni of faint SNe\,Ia --- for SN1991bg, \citet{c+97} model the V-band light curve and 
obtain a mass of 0.1 M$_{\odot}$, and \citet{m+97} model the photospheric and nebular-epoch 
spectra and obtain $^{56}$Ni mass of 0.07 M$_{\odot}$.

\section{Discussion}

To summarize, the primary observational characteristics of subluminous SNe\,Ia 
(of which SN2007ax is an extreme case) are small t$_{\rm b}$ in the
optical $B$-band light curve, extremely red $B-V$ color at maximum, possible excess in UV
emission at late-time, presence of intermediate mass elements in spectra,
medium ejecta velocities, low $^{56}$Ni mass in ejecta and short rise-times. 

Several theoretical models have been proposed to explain faint SNe\,Ia 
-- complete detonation of a sub-Chandrashekhar
mass white dwarf, a delayed detonation model, a failed neutron star model 
and a small scale deflagration model. The detonation of a sub-Chandrashekhar 
C-O white dwarf (e.g. \citealt{l90}, \citealt{ww94}) produces more $^{56}$Ni than observed 
and is more blue at maximum than observed \citep{hk96}. If we consider detonation of 
a sub-Chandrashekhar O-Ne-Mg white dwarf \citep{i+91}, the total nuclear
energy is smaller and the predicted ejecta velocities are lower than observed \citep{f+92}. 
\citet{m+07} use detailed spectral modeling to show a common explosion mechanism for all SNe\,Ia, 
likely delayed detonation. The failed neutron star model \citep{ni85} suggests that 
if the accretion rate of carbon and oxygen from a companion onto a white dwarf is high enough, 
it may prematurely ignite CO on the white dwarf surface. Thus, instead of a neutron star, we may 
see a faint SNe\,Ia. Small-scale deflagration models suggest that either the burning is restricted 
to the outer layers or that it occurs slowly.

Another intriguing theoretical possibility recently proposed by \citet{b+07} is 
faint thermonuclear supernovae from ultracompact double degenerate AM CVn systems. 
This supernova is tantalizingly at the brightest end of their predictions 
(M$_{V}$=$-$14 to $-$16, timescale = 2--6 days, M$_{ej} <$ 0.1 M$\odot$).
However, the decay time predicted by these models is much shorter and the $^{56}$Ni mass
less than that observed in SN2007ax. Also, the spectrum does not show any feature 
which suggests being powered by different radioactive material ($^{48}$Cr, $^{44}$Ti, $^{52}$Fe) 
produced by some of these models. 

None of the above models convincingly explain all the observed characteristics 
of subluminous SNe\,Ia. SN2007ax compels the question of what is the (and whether there is) 
lower limit of $^{56}$Ni mass in a thermonuclear explosion. 
Only if we can explain the extremely faint (and the extremely bright) supernovae
will we thoroughly understand the limitations in physical processes 
involved in the thermonuclear explosion, in particular, the 
$^{56}$Ni mass production. 

Future supernova surveys which have a shorter cadence and a  deeper limiting magnitude will 
provide invaluable clues to understanding the nature of subluminous SNe\,Ia. Follow-up of
these supernovae with well-sampled UV light curves and well-calibrated multi-epoch UV
spectra would also be important to understand the apparent excess at late-time.

We suggest how a near-future survey, for example, the Palomar Transient Factory\footnote{The Palomar 
Transient Factory is a dedicated time-domain astronomy project to come online on Palomar 48-inch in Nov 08.}, 
can systematically search for faint SNe\,Ia. The parameters of the survey design are sky coverage, cadence, depth,
filter and choice of pointings. \citet{h01} shows that faint SNe\,Ia occur 
preferentially in early-type galaxies and \citet{t+07} suggest that they occur in lower metallicity, 
old stellar mass populations. Since they decline by a magnitude in five days, the cadence of the
search should be faster than five days so that the detection sample is complete. Since faint SNe\,Ia are 
extremely red at maximum, we should choose a red filter for the search. To maximize sky coverage,
searching with a single red filter should suffice (with multi-band follow-up).
Since the local universe is clumpy (e.g. : $\approx$25\% of the total light at the distance of Virgo is
in the Virgo supercluster), the sky coverage must include concentrations in stellar mass, such as 
the Virgo, Perseus and Coma galaxy clusters. The rate of normal SNe\,Ia is 3 per 
10$^{11}$L$_{\odot}$ per century \citep{sb05}. \citet{l+01} estimate a rate for subluminous
SNe\,Ia to be 16\% of normal SNe\,Ia rate based on LOSS and BAOSS surveys.  To a depth of absolute magnitude 
of $-$15.5, and with a limiting magnitude of 20.5, the survey volume would be 1.5$\times$ 10$^{7}$ Mpc$^3$. Using 
the 2MASS K-band luminosity function of 
5.1$\times $10$^{8} $L$_{\odot}$Mpc$^{-3}$ (\citealt{kk05},~\citealt{k+01}), we expect a rate of the 
faintest subluminous supernovae to be $\approx$370 all sky per year. The Palomar Transient Factory
plans a 5-day cadence 2700 sq. deg. experiment which would give $\approx$24 faint SNe\,Ia per year.

\acknowledgements

We thank Nick Scoville, Milan Bogoslavejic and the Swift team for performing our 
Target of Opportunity observations flawlessly. We would like to thank Brent Tully 
for providing his catalog of nearby galaxies. LB thanks NSF grants PHY 05-51164 and AST 02-05956. 

\facility{{\it Facilities:} \facility{PO:1.5}, \facility{Hale (DBSP)}, \facility{Keck:I (LRIS)}, \facility{Keck:II (NIRC2)}, \facility{Swift (UVOT)}, \facility{Bok}, \facility{SLOTIS}}










\clearpage

\begin{center}
\begin{longtable}{llllll}
\caption[]{\bf Summary of Photometric Observations of SN2007ax } \cr 
\hline
\hline
UT & MJD & Facility & Band & Exp & Magnitude \cr
\hline 

\hline
04-Apr-2007 & 54194.29 & Keck-NIRC2   & K' & 600\,s & $16.7\pm 0.1$\,mag\cr

\hline
29-Mar-2007 & 54188.147 & Palomar-60 & $r'$ & $3\times 180\,$s & 15.65 $\pm$ 0.06\,mag\cr 
30-Mar-2007 & 54189.230 & Palomar-60 & $r'$ & $3\times 180\,$s & 15.75 $\pm$ 0.05\,mag\cr
30-Mar-2007 & 54189.331 & Palomar-60 & $r'$ & $3\times 180\,$s & 15.60 $\pm$ 0.02\,mag\cr
31-Mar-2007 & 54190.232 & Palomar-60 & $r'$ & $3\times 180\,$s & 15.73 $\pm$ 0.03\,mag\cr
31-Mar-2007 & 54190.329 & Palomar-60 & $r'$ & $3\times 180\,$s & 15.59 $\pm$ 0.03\,mag\cr
01-Apr-2007 & 54191.137 & Palomar-60 & $r'$ & $3\times 180\,$s & 15.68 $\pm$ 0.05\,mag\cr
01-Apr-2007 & 54191.234 & Palomar-60 & $r'$ & $3\times 180\,$s & 15.69 $\pm$ 0.03\,mag\cr
01-Apr-2007 & 54191.137 & Palomar-60 & $r'$ & $3\times 180\,$s & 15.68 $\pm$ 0.05\,mag\cr
02-Apr-2007 & 54192.233 & Palomar-60 & $r'$ & $3\times 180\,$s & 15.36 $\pm$ 0.04\,mag\cr
03-Apr-2007 & 54193.138 & Palomar-60 & $r'$ & $3\times 180\,$s & 15.56 $\pm$ 0.03\,mag\cr
03-Apr-2007 & 54193.231 & Palomar-60 & $r'$ & $3\times 180\,$s & 15.59 $\pm$ 0.22\,mag\cr
04-Apr-2007 & 54194.238 & Palomar-60 & $r'$ & $3\times 180\,$s & 15.45 $\pm$ 0.04\,mag\cr
05-Apr-2007 & 54195.223 & Palomar-60 & $r'$ & $3\times 180\,$s & 15.43 $\pm$ 0.03\,mag\cr
06-Apr-2007 & 54196.239 & Palomar-60 & $r'$ & $3\times 180\,$s & 15.50 $\pm$ 0.03\,mag\cr
07-Apr-2007 & 54197.307 & Palomar-60 & $r'$ & $3\times 180\,$s & 15.94 $\pm$ 0.06\,mag\cr
11-Apr-2007 & 54201.145 & Palomar-60 & $r'$ & $3\times 180\,$s & 16.41 $\pm$ 0.07\,mag\cr
14-Apr-2007 & 54204.146 & Palomar-60 & $r'$ & $3\times 180\,$s & 16.69 $\pm$ 0.03\,mag\cr
14-Apr-2007 & 54204.245 & Palomar-60 & $r'$ & $3\times 180\,$s & 16.31 $\pm$ 0.12\,mag\cr
18-Apr-2007 & 54208.165 & Palomar-60 & $r'$ & $3\times 180\,$s & 17.09 $\pm$ 0.04\,mag\cr
25-Apr-2007 & 54215.164 & Palomar-60 & $r'$ & $3\times 180\,$s & 17.58 $\pm$ 0.07\,mag\cr
28-Apr-2007 & 54218.157 & Palomar-60 & $r'$ & $3\times 180\,$s & 17.62 $\pm$ 0.05\,mag\cr
29-Apr-2007 & 54219.195 & Palomar-60 & $r'$ & $3\times 180\,$s & 17.64 $\pm$ 0.14\,mag\cr
30-Apr-2007 & 54220.154 & Palomar-60 & $r'$ & $3\times 180\,$s & 17.74 $\pm$ 0.09\,mag\cr
08-May-2007 & 54228.159 & Palomar-60 & $r'$ & $3\times 180\,$s & 18.18 $\pm$ 0.18\,mag\cr

\hline
29-Mar-2007 & 54188.127 & Palomar-60 & $g'$ & $3\times 180\,$s & 16.27 $\pm$ 0.09\,mag\cr
29-Mar-2007 & 54188.142 & Palomar-60 & $g'$ & $3\times 180\,$s & 16.30 $\pm$ 0.05\,mag\cr
30-Mar-2007 & 54189.128 & Palomar-60 & $g'$ & $3\times 180\,$s & 16.37 $\pm$ 0.04\,mag\cr
30-Mar-2007 & 54189.222 & Palomar-60 & $g'$ & $3\times 180\,$s & 16.23 $\pm$ 0.11\,mag\cr
30-Mar-2007 & 54189.323 & Palomar-60 & $g'$ & $3\times 180\,$s & 16.20 $\pm$ 0.14\,mag\cr
31-Mar-2007 & 54190.129 & Palomar-60 & $g'$ & $3\times 180\,$s & 16.36 $\pm$ 0.08\,mag\cr
31-Mar-2007 & 54190.225 & Palomar-60 & $g'$ & $3\times 180\,$s & 16.37 $\pm$ 0.12\,mag\cr
31-Mar-2007 & 54190.321 & Palomar-60 & $g'$ & $3\times 180\,$s & 16.22 $\pm$ 0.12\,mag\cr
01-Apr-2007 & 54191.130 & Palomar-60 & $g'$ & $3\times 180\,$s & 16.45 $\pm$ 0.11\,mag\cr
01-Apr-2007 & 54191.227 & Palomar-60 & $g'$ & $3\times 180\,$s & 16.43 $\pm$ 0.12\,mag\cr
01-Apr-2007 & 54191.130 & Palomar-60 & $g'$ & $3\times 180\,$s & 16.30 $\pm$ 0.13\,mag\cr
02-Apr-2007 & 54192.130 & Palomar-60 & $g'$ & $3\times 180\,$s & 16.49 $\pm$ 0.11\,mag\cr
02-Apr-2007 & 54192.226 & Palomar-60 & $g'$ & $3\times 180\,$s & 16.39 $\pm$ 0.12\,mag\cr
02-Apr-2007 & 54192.322 & Palomar-60 & $g'$ & $3\times 180\,$s & 16.39 $\pm$ 0.09\,mag\cr
03-Apr-2007 & 54193.224 & Palomar-60 & $g'$ & $3\times 180\,$s & 16.45 $\pm$ 0.11\,mag\cr
03-Apr-2007 & 54193.320 & Palomar-60 & $g'$ & $3\times 180\,$s & 16.53 $\pm$ 0.12\,mag\cr
04-Apr-2007 & 54194.132 & Palomar-60 & $g'$ & $3\times 180\,$s & 16.59 $\pm$ 0.09\,mag\cr
04-Apr-2007 & 54194.230 & Palomar-60 & $g'$ & $3\times 180\,$s & 16.26 $\pm$ 0.11\,mag\cr
05-Apr-2007 & 54195.216 & Palomar-60 & $g'$ & $3\times 180\,$s & 16.47 $\pm$ 0.09\,mag\cr
06-Apr-2007 & 54196.138 & Palomar-60 & $g'$ & $3\times 180\,$s & 16.66 $\pm$ 0.07\,mag\cr
06-Apr-2007 & 54196.231 & Palomar-60 & $g'$ & $3\times 180\,$s & 17.06 $\pm$ 0.12\,mag\cr
07-Apr-2007 & 54197.300 & Palomar-60 & $g'$ & $3\times 180\,$s & 17.03 $\pm$ 0.07\,mag\cr
11-Apr-2007 & 54201.138 & Palomar-60 & $g'$ & $3\times 180\,$s & 17.62 $\pm$ 0.14\,mag\cr
11-Apr-2007 & 54201.235 & Palomar-60 & $g'$ & $3\times 180\,$s & 17.47 $\pm$ 0.17\,mag\cr
14-Apr-2007 & 54204.139 & Palomar-60 & $g'$ & $3\times 180\,$s & 17.80 $\pm$ 0.13\,mag\cr
14-Apr-2007 & 54204.237 & Palomar-60 & $g'$ & $3\times 180\,$s & 17.59 $\pm$ 0.12\,mag\cr
17-Apr-2007 & 54207.159 & Palomar-60 & $g'$ & $3\times 180\,$s & 18.11 $\pm$ 0.19\,mag\cr
18-Apr-2007 & 54208.158 & Palomar-60 & $g'$ & $3\times 180\,$s & 18.04 $\pm$ 0.13\,mag\cr
18-Apr-2007 & 54208.275 & Palomar-60 & $g'$ & $3\times 180\,$s & 17.98 $\pm$ 0.15\,mag\cr
20-Apr-2007 & 54210.140 & Palomar-60 & $g'$ & $3\times 180\,$s & 18.09 $\pm$ 0.23\,mag\cr
22-Apr-2007 & 54212.267 & Palomar-60 & $g'$ & $3\times 180\,$s & 18.20 $\pm$ 0.23\,mag\cr
25-Apr-2007 & 54215.157 & Palomar-60 & $g'$ & $3\times 180\,$s & 18.49 $\pm$ 0.42\,mag\cr
26-Apr-2007 & 54216.155 & Palomar-60 & $g'$ & $3\times 180\,$s & 18.61 $\pm$ 0.34\,mag\cr
28-Apr-2007 & 54218.150 & Palomar-60 & $g'$ & $3\times 180\,$s & 18.57 $\pm$ 0.39\,mag\cr
29-Apr-2007 & 54219.188 & Palomar-60 & $g'$ & $3\times 180\,$s & 18.42 $\pm$ 0.22\,mag\cr
30-Apr-2007 & 54220.147 & Palomar-60 & $g'$ & $3\times 180\,$s & 18.73 $\pm$ 0.40\,mag\cr
09-May-2007 & 54229.159 & Palomar-60 & $g'$ & $3\times 180\,$s & 18.71 $\pm$ 0.10\,mag\cr

\hline
02-Apr-2007 & 54192.8 & Swift-UVOT & $v$ &   419\,s  & 15.87 $\pm$ 0.08\,mag\cr
04-Apr-2007 & 54195.0 & Swift-UVOT & $v$ &   229\,s  & 16.10 $\pm$ 0.09\,mag\cr   
10-Apr-2007 & 54201.0 & Swift-UVOT & $v$ &   227\,s  & 16.75 $\pm$ 0.14\,mag\cr
17-Apr-2007 & 54207.2 & Swift-UVOT & $v$ &   225\,s  & 17.42 $\pm$ 0.23\,mag\cr
21-Apr-2007 & 54211.4 & Swift-UVOT & $v$ &   516\,s  & 17.94 $\pm$ 0.33\,mag\cr

\hline
29-Mar-2007 & 54189.0 & Swift-UVOT & $b$ &   189\,s  & 16.96 $\pm$ 0.05\,mag\cr
02-Apr-2007 & 54192.8 & Swift-UVOT & $b$ &   419\,s  & 17.22 $\pm$ 0.05\,mag\cr
04-Apr-2007 & 54195.0 & Swift-UVOT & $b$ &   205\,s  & 18.02 $\pm$ 0.11\,mag\cr   
10-Apr-2007 & 54201.0 & Swift-UVOT & $b$ &   202\,s  & 18.57 $\pm$ 0.16\,mag\cr
17-Apr-2007 & 54207.2 & Swift-UVOT & $b$ &   324\,s  & 18.62 $\pm$ 0.14\,mag\cr
21-Apr-2007 & 54211.4 & Swift-UVOT & $b$ &   679\,s  & 18.64 $\pm$ 0.12\,mag\cr
30-Apr-2007 & 54220.2 & Swift-UVOT & $b$ &   677\,s  & 19.70 $\pm$ 0.29\,mag\cr
06-May-2007 & 54226.3 & Swift-UVOT & $b$ &   134\,s  & 19.15 $\pm$ 0.30\,mag\cr
11-May-2007 & 54231.9 & Swift-UVOT & $b$ &   723\,s  & $ > 19.8$\,mag\cr
              
\hline
02-Apr-2007 & 54192.8 & Swift-UVOT & $u$ &  419\,s   &  17.82 $\pm$ 0.08\,mag\cr
04-Apr-2007 & 54194.0 & Swift-UVOT & $u$ &  229\,s   &  18.26 $\pm$ 0.13\,mag \cr
10-Apr-2007 & 54201.0 & Swift-UVOT & $u$ &  227\,s   &  18.94 $\pm$ 0.22\,mag\cr
17-Apr-2007 & 54207.2 & Swift-UVOT & $u$ &  648\,s   &  18.86 $\pm$ 0.14\,mag\cr

\hline
29-Mar-2007 & 54188.9 & Swift-UVOT & $uvw1$ &  432\,s   & 20.02\footnote[a]{The significance of this detection is 2.5$\sigma$} $\pm$ 0.47\,mag\cr
02-Apr-2007 & 54192.8 & Swift-UVOT & $uvw1$ &  844\,s   & 20.07 $\pm$ 0.40\,mag\cr
04-Apr-2007 & 54195.2 & Swift-UVOT & $uvw1$ &  1669\,s  & $ > 20.42$\,mag\cr
10-Apr-2007 & 54201.2 & Swift-UVOT & $uvw1$ &  1369\,s  & $ > 20.45$\,mag\cr
17-Apr-2007 & 54207.2 & Swift-UVOT & $uvw1$ &  1620\,s  & $ > 20.42$\,mag\cr
21-Apr-2007 & 54211.4 & Swift-UVOT & $uvw1$ &  3400\,s  & 20.51  $\pm$ 0.40 \,mag\cr
30-Apr-2007 & 54220.7 & Swift-UVOT & $uvw1$ &  3390\,s  & 20.86\footnote[b]{The significance of this detection is 2.3$\sigma$}  $\pm$ 0.50 \,mag\cr
06-May-2007 & 54226.7 & Swift-UVOT & $uvw1$ &  6681\,s  & $ > 20.12$\,mag\cr
11-May-2007 & 54232.4 & Swift-UVOT & $uvw1$ &  3615\,s  & $ > 20.62$\,mag\cr

\hline
18-Apr-2007 & 54208.6 & Super-LOTIS & $I$ & 13 $\times$ 60\,s & 17.01 $\pm$ 0.02 \,mag\cr
19-Apr-2007 & 54209.6 & Super-LOTIS & $I$ & 10 $\times$ 60\,s & 17.22 $\pm$ 0.02 \,mag\cr
20-Apr-2007 & 54210.6 & Super-LOTIS & $I$ & 9 $\times$ 60\,s & 17.18 $\pm$ 0.02 \,mag\cr
27-Apr-2007 & 54217.6 & Super-LOTIS & $I$ & 6 $\times$ 60\,s & 17.59 $\pm$ 0.04 \,mag\cr
04-May-2007 & 54224.6 & Super-LOTIS & $I$ & 18 $\times$ 60\,s & 18.00 $\pm$ 0.05 \,mag\cr
11-May-2007 & 54231.6 & Super-LOTIS & $I$ & 16 $\times$ 60\,s & 18.35 $\pm$ 0.10 \,mag\cr
26-May-2007 & 54246.6 & Super-LOTIS & $I$ & 18 $\times$ 60\,s & 18.90 $\pm$ 0.29 \,mag\cr
29-May-2007 & 54249.6 & Super-LOTIS & $I$ & 22 $\times$ 60\,s & 19.05 $\pm$ 0.18 \,mag\cr
30-May-2007 & 54250.6 & Super-LOTIS & $I$ & 14 $\times$ 60\,s & 19.15 $\pm$ 0.10 \,mag\cr

\hline
18-Apr-2007 & 54208.6 & Super-LOTIS & $R$ & 12 $\times$ 60\,s & 16.79 $\pm$ 0.23 \,mag\cr
19-Apr-2007 & 54209.6 & Super-LOTIS & $R$ & 12 $\times$ 60\,s & 16.97 $\pm$ 0.22 \,mag\cr
20-Apr-2007 & 54210.6 & Super-LOTIS & $R$ & 14 $\times$ 60\,s & 17.10 $\pm$ 0.20 \,mag\cr
23-Apr-2007 & 54213.6 & Super-LOTIS & $R$ & 15 $\times$ 60\,s & 17.24 $\pm$ 0.22 \,mag\cr
26-Apr-2007 & 54216.6 & Super-LOTIS & $R$ & 47 $\times$ 60\,s & 17.32 $\pm$ 0.29 \,mag\cr
27-Apr-2007 & 54217.6 & Super-LOTIS & $R$ & 12 $\times$ 60\,s & 17.47 $\pm$ 0.29 \,mag\cr
04-May-2007 & 54224.6 & Super-LOTIS & $R$ & 10 $\times$ 60\,s & 17.72 $\pm$ 0.38 \,mag\cr
11-May-2007 & 54231.6 & Super-LOTIS & $R$ & 16 $\times$ 60\,s & 17.96 $\pm$ 0.46 \,mag\cr
26-May-2007 & 54246.6 & Super-LOTIS & $R$ & 17 $\times$ 60\,s & 18.11 $\pm$ 0.40 \,mag\cr
27-May-2007 & 54247.6 & Super-LOTIS & $R$ & 11 $\times$ 60\,s & 18.55 $\pm$ 0.37 \,mag\cr
31-May-2007 & 54251.6 & Super-LOTIS & $R$ & 11 $\times$ 60\,s & 18.16 $\pm$ 0.46 \,mag\cr
05-Jun-2007 & 54256.6 & Super-LOTIS & $R$ & 20 $\times$ 60\,s & 18.20 $\pm$ 0.54 \,mag\cr

\hline
18-Apr-2007 & 54208.6 & Super-LOTIS & $V$ & 9 $\times$ 60\,s & 17.84 $\pm$ 0.15 \,mag\cr
19-Apr-2007 & 54209.6 & Super-LOTIS & $V$ & 6 $\times$ 60\,s & 17.80 $\pm$ 0.15 \,mag\cr
20-Apr-2007 & 54210.6 & Super-LOTIS & $V$ & 8 $\times$ 60\,s & 17.80 $\pm$ 0.15 \,mag\cr
27-Apr-2007 & 54217.6 & Super-LOTIS & $V$ & 16 $\times$ 60\,s & 18.30 $\pm$ 0.15 \,mag\cr
04-May-2007 & 54224.6 & Super-LOTIS & $V$ & 9 $\times$ 60\,s & 18.94 $\pm$ 0.15 \,mag\cr
11-May-2007 & 54231.6 & Super-LOTIS & $V$ & 17 $\times$ 60\,s & 18.87 $\pm$ 0.15 \,mag\cr
27-May-2007 & 54247.6 & Super-LOTIS & $V$ & 18 $\times$ 60\,s & 19.55 $\pm$ 0.23 \,mag\cr
29-May-2007 & 54249.6 & Super-LOTIS & $V$ & 22 $\times$ 60\,s & 19.27 $\pm$ 0.16 \,mag\cr

\hline
27-May-2007 & 54247.6 & Bok-2.3m & $B$ & 4 $\times$ 180\,s &    20.53 $\pm$ 0.04 \,mag\cr

\hline
\hline

\label{tab:obs}
\end{longtable}
\end{center}

\end{document}